# Raspberry PI for compact autonomous home farm control

Ildar Rakhmatulin, PhD

Department of Power Plants Networks and Systems, South Ural State University
ildar.o2010@yandex.ru

**Abstract**
This manuscript presented an autonomous home farm for predicting metrological characteristics that can not only automate the process of growing crops but also, due to a neural network, significantly increase the productivity of the farm. The developed farm monitors and manages the following indicators: illumination, soil PH, air temperature, ground temperature, air humidity, $CO_2$ concentration, and soil moisture. The presented farm can also be considered as a device for testing various weather conditions to determine the optimal temperature characteristics for different crops. This farm as a result is completely autonomous grows tomatoes at home.

**Introduction**
Every year the population in the planet grows, but the size of the cultivated area decreases. Colezea, et al. (1, 2018) and Heike, et al. (2, 2018) predicted food shortages in the following decades. The solutions to this problem are to increase the productivity of the fields using herbicides (3, 2015; 4, 2008) and controlling weeds with higher effectively (5, 2021; 6, 2020). We suggest another way - an autonomous home farm. Today, due to several shortcomings, home farms are not widely used in large cities, since a lot of time must be spent on caring for the crop. But due to neural networks, the farm can not only be automated but also increase its efficiency by creating an ideal microclimate for each crop grown.

**Materials and method**
We have created a small home farm - with the ability to control and create a microclimate. Measuring and controlling the following indicators:

- Illumination;
- PH ground;
- Air temperature;
- Ground temperature;
- Air humidity;
- $CO_2$ concentration;
- Soil moisture.

Ultimately, through machine learning, we taught our farm to automatically adjust the parameters of the devices to create the ideal microclimate.

The operation algorithm of the device is shown in Fig. 1.

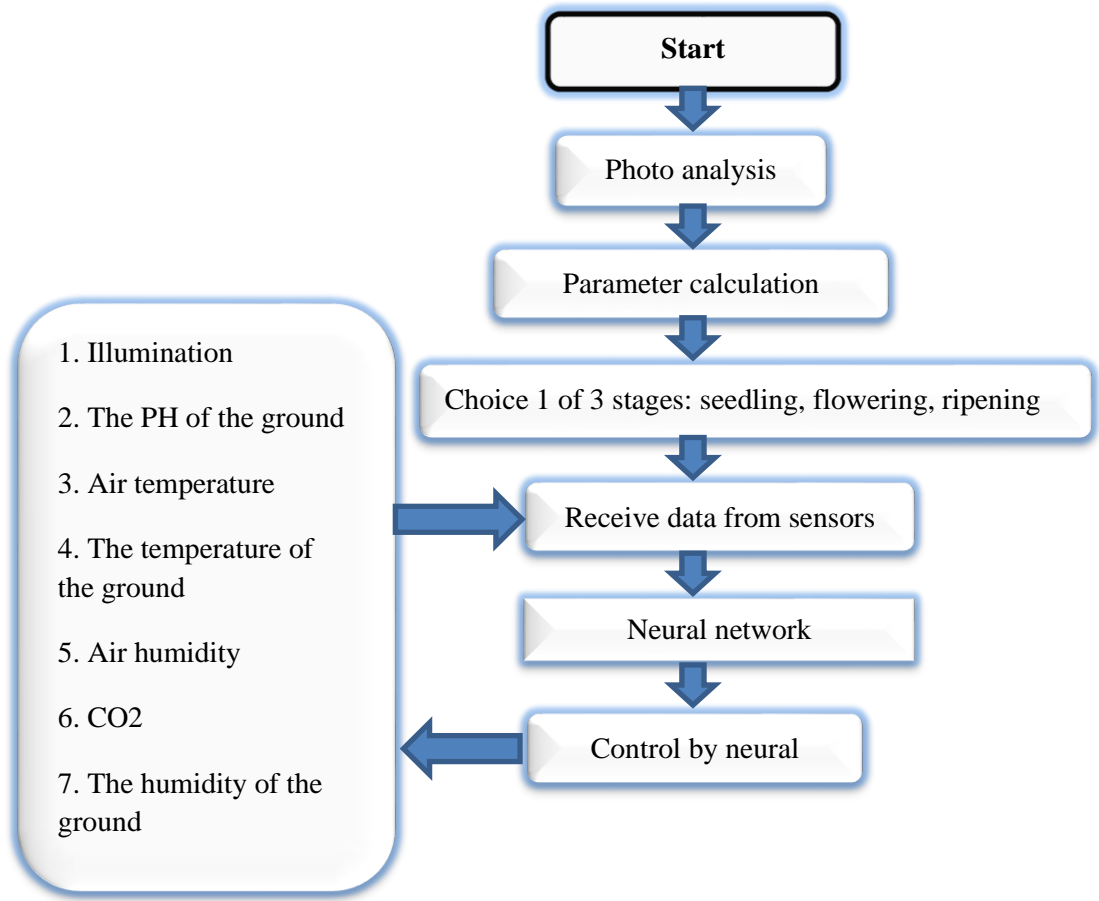

Fig. 1 Algorithm of the home farm

To create a microclimate, we used the following equipment:

- $CO_2$ - controlled by a digital sensor – ccs811. We can only reduce a value of $CO_2$ in farm by fan which delivers fresh air. (One fan yet we used only during flowering for pollination):

- Temperature and humidity measured by a DHT 11 sensor. Calibration by the Metre USB Data Logger GM1361 device. The temperature controlled by ceramic electric heating plates;

- The soil temperature measured by pt100, for ground heating used ceramic electric heating plates;

- The moisture of the soil controlled by an analog device calibrated on the VBESTLIFE di-vital Soil Humidity Sensor. For irrigate used a dosing pump.

- Raspberry Pi 3, Broadcom BCM2837B0 (ARM Cortex-A53);

- Software - Python 3.6, OpenCV 3.4.1, PyQt5;

- Camera, Sony IMX219 Exmor;

- Hilitand SM206 digital solar meter to measure the light emission. The brightness of the light is regulated by means of pulse-width modulation on a phytolamp with LEDs with a wavelength of 620 nm ~ 750 nm;

- opt101 sensor to measure the intensity of solar radiation (calibrated by Sun Light Radiation Measuring Testing Instrument, Hilitand SM206 Digital Solar Metre).

Figure 2 shows a schematic diagram of the created farm.

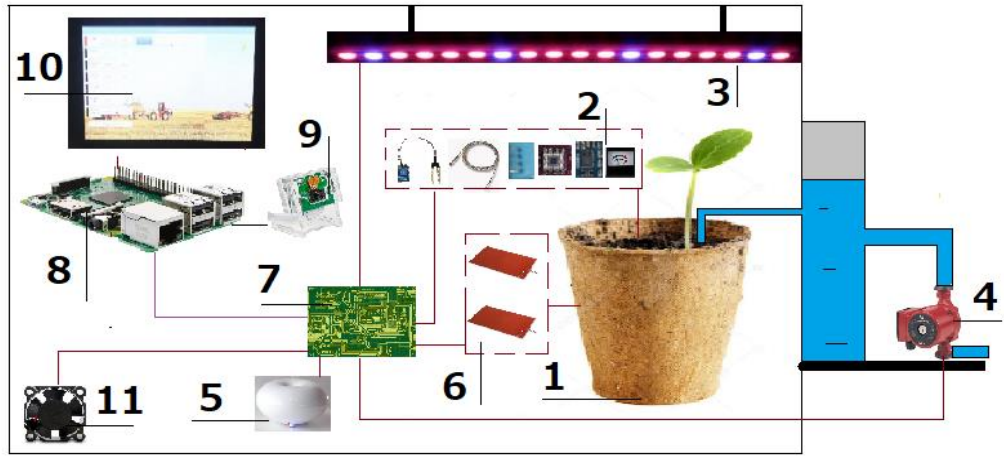

Fig.2. The farm operation scheme: 1 - crop, 2 - sensors, 3 - LED, 4 - pump, 5 - humidifier, 6 - air and soil heaters, 7 – analog digital converter, 8 - RaspberryPI, 9 - camera, 10 - operator panel, 11- fan.

**3. Experience**

We simulated a wide range of different situations in which different sensors and actuators were involved. We have artificially changed the parameters to explore the possibility of finding the right compensation to increase the yield of the cultivated crop. For example, we lowered the illumination of a tomato during the germination period to 3500 lux, and then monitored the readings from all sensors. After that, we tried to compensate for the lack of color with different parameters. As a result, independent data were collected. Fig. 3 shows the electrical part of the farm.

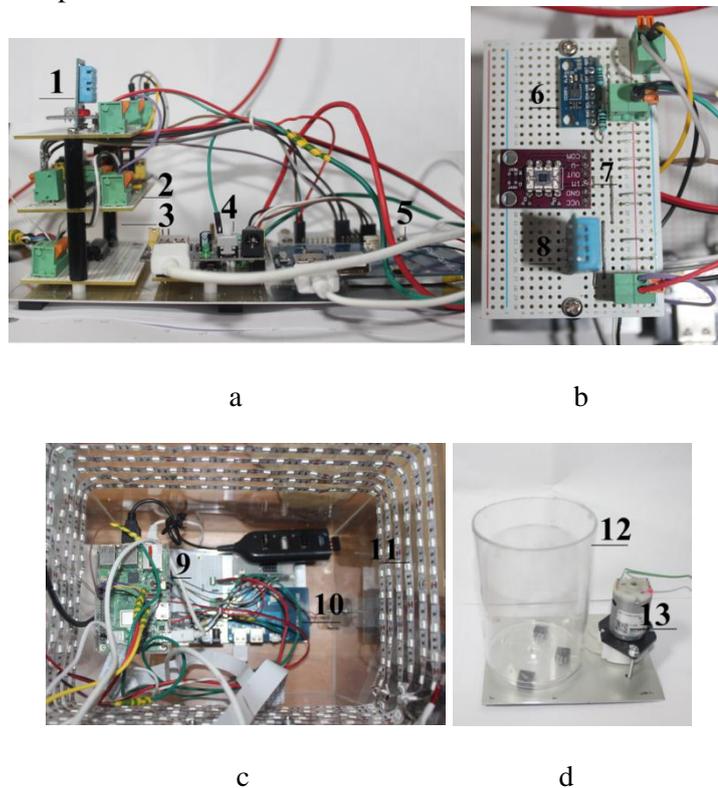

a  b

c  d

Fig. 3. Electrical part farm 1 - board with sensors, 2 ,3 - analog boards for signals processing, 4 - power board, 5 - battery, 6 - CO2 sensor, 7 - illumination sensor, 8 - temperature sensor, 9 - RaspberryPI3, 10 - battery board, 11 - LED, 12 - vessel for water, 13 - pump

Research in this area has two options for implementation: data processing directly on one single-board computer in our case, Raspberry PI, or data transfer to the cloud and visualization via a web application. We developed software that can help grow a variety of fruits and vegetables. The program is written in Python 3.6 (PyQt5 library). The user menu contains many settings and functions for monitoring the conditions of the farm. In the menu is possible to view the parameters of the neural network, find out the forecast of product availability, and view online graphs and the state of control mechanisms. The user interface is controlled on the touch screen, the general view of the interface is shown in Fig. 4.

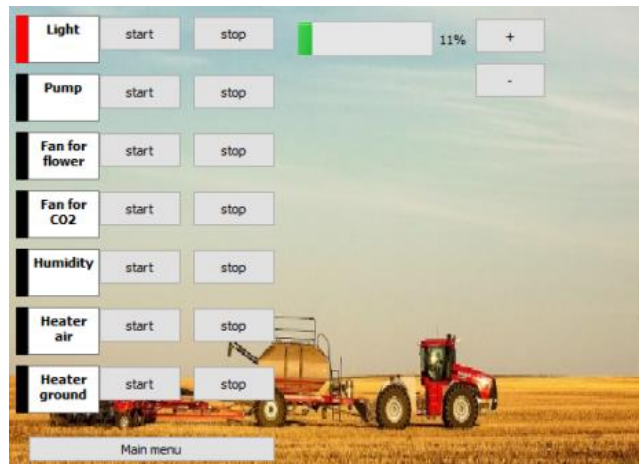

Fig. 4. Appearance of the user interface

The sensor reading is slightly unstable. But this does not affect the result and is explained by the noise of the sensors, which, if necessary, can be replaced by sensors with a higher resolution.
General view of the farm, in Fig. 5.

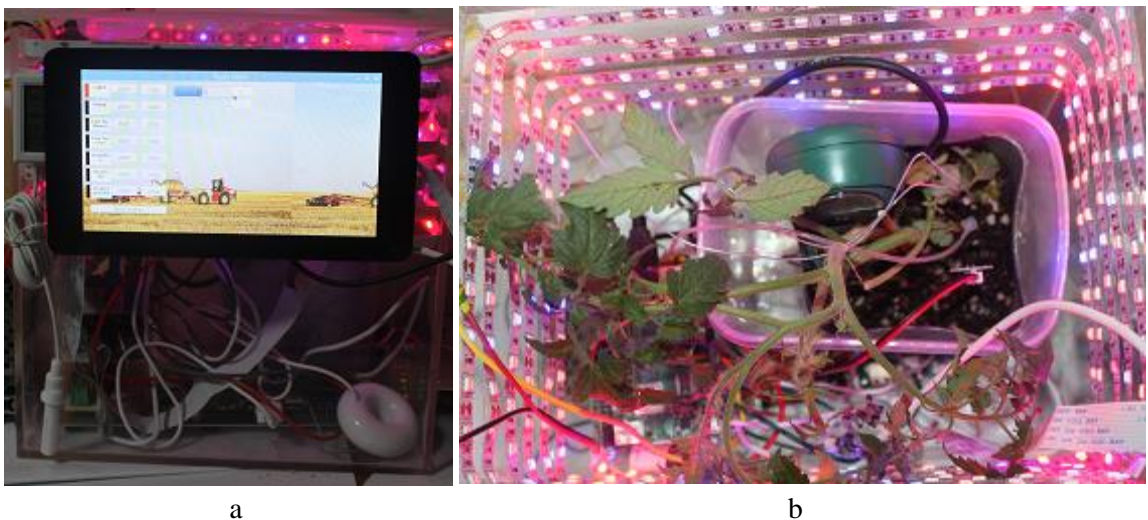

a              b
Fig.5. General View of the farm: a - side view, b - view from above

Several works have similar topics, but they consider a rather narrow range of influencing factors, so the results obtained in these works cannot be used in this manuscript. Presented farms in these papers have common problems as high price, difficulty in maintenance, and frequent breakdown (7, 2019; 8, 2011; 9, 2021).
In paper (10, 2021), we made an overview of neural networks that were used in agriculture. Based on what we created neural network model. As input data, we have signals from 8 sensors. The neural network can find the optimal equations for the entire set of sensors and trained the system to compensate for external weather factors.

Ambient temperature is one of the main indicators for both agriculture and measurement equipment. At a stable temperature, the error of all sensors used on the farm is presented in the technical documentation and does not exceed 2%. Temperature changes during farm operation are critical to any analog measurement equipment. The sensors are in the same ecosystem as farming on a farm, which means they are exposed to the same temperature extremes. With a linear decrease in temperature, the measurement becomes unstable, and there is no direct correlation between temperature and measurement error. The radiation intensity sensor has a small error due to the technical design of the sensor, for example, the absence of ceramic capacitors, etc. But in the sensor for measuring PH, the measurement results change depending on temperature by more than 10% percent, which is a critical value for mathematical calculations. To solve this problem, we trained the neural network at different ambient temperatures. The neural network received information about the temperature in the ecosystem. Therefore, with a large amount of data, the network can learn to determine the nature of the influence of temperature factors on the magnitude of the error of the measuring equipment. In the future, it is necessary to conduct studies in which it will be clear how much data is needed for the neural network for it to compensate for the measurement error of the sensors. In a trained network, there may be problems with replacing sensors, since in this case the parameters of the sensor change, and it will take time to retrain the network.

**Discussion and conclusion**
As a result, we got a fully autonomous low-cost farm with the ability to predict the yield and control executive mechanisms to control the ecosystem of the farm.
In this paper, we did not consider many factors as nitrogen content in soil and leaves, irrigation, determination of water deficit in plants, assessment of water erosion, detection of pests, the use of herbicides, detection of contaminants, diseases, or food defects. Only one type of landing was considered. Possibly consider the following growing media - aerobic, aquaporin, or hydroponic nutrient. Therefore, the research carried out in this article can be supplemented and continued with absolutely any type.